\documentclass[twocolumn]{aastex701}

\usepackage[colorinlistoftodos]{todonotes}
\usepackage{placeins}
\usepackage{hyperref}
\usepackage{amssymb}
\usepackage{comment}
\usepackage{soul}
\usepackage[export]{adjustbox}
\usepackage{relsize}

\definecolor{colorTC}{rgb}{.2,.7,.2}

\usepackage{array}
\usepackage{amsmath}
\newcolumntype{P}[1]{>{\centering\arraybackslash}p{#1}}
\newcolumntype{R}[1]{>{\raggedleft\arraybackslash}p{#1}}
\newcolumntype{L}[1]{>{\raggedright\arraybackslash}p{#1}}
\usepackage{longtable}
\usepackage[font=small,labelfont=bf]{caption} % optional styling
\renewcommand{\j}{\ensuremath{\text{j}}}

\newcommand{\s}{\hspace{0.8pt}}
\newcommand{\secspacing}{\vspace{10.pt}}
\newcommand{\cg}[1]{\boldsymbol{#1}}
\newcommand{\vcg}[1]{\vec{\boldsymbol{#1}}}
\let\v\relax
\newcommand{\v}[1]{\vec{#1}}
\newcommand{\diff}{\text{d}}
\newcommand{\mdm}{m_{\rm dm}}
\newcommand{\adm}{\alpha_{\rm dm}}
\newcommand{\mm}{m_{V}}
\newcommand{\g}{v_{\rm rel}}
\newcommand{\G}{V}

\usepackage{cleveref}
\crefname{table}{Table}{Tables}
\crefname{equation}{Eq.}{Eqs.}
\crefname{appendix}{App.}{Apps.}
\crefname{section}{Sec.}{Secs.}
\crefname{figure}{Fig.}{Figs.}

\makeatletter
\g@addto@macro\bfseries{\boldmath}
\makeatother

\begin{document}

{\raggedleft CERN-TH-2025-263\par}
\vskip -10 pt

\title{First-Principles Formalism for Simulating Self-Interacting Dark Matter}

\author[orcid=0000-0001-7743-7364,gname=Maria,sname=Ramos]{Maria Ramos}
\affiliation{Theoretical Physics Department, CERN, 1211 Geneva, Switzerland}
\email[show]{maria.ramos@cern.ch}  

\author[orcid=0000-0002-7040-3038,gname=Timothy,sname=Cohen]{Timothy Cohen}
\affiliation{Theoretical Physics Department, CERN, 1211 Geneva, Switzerland}
\affiliation{Theoretical Particle Physics Laboratory, EPFL, 1015 Lausanne, Switzerland}
\affiliation{Institute for Fundamental Science, University of Oregon, Eugene, OR 97403, USA}
\email[hide]{tim.cohen@cern.ch}  

\author[orcid=0000-0002-8495-8659,gname=Mariangela,sname=Lisanti]{Mariangela Lisanti\,}
\affiliation{Department of Physics, Princeton University, Princeton, NJ 08544, USA}
\affiliation{Center for Computational Astrophysics, Flatiron Institute, New York, NY 10010, USA}
\email[hide]{mlisanti@princeton.edu}

\begin{abstract}

\noindent It is plausible that the dark matter particles have non-gravitational interactions among themselves.  If such self interactions are large enough, they could leave an imprint on the morphology of galaxies.  These effects can be studied with numerical simulations, which serve as the primary tool to predict the non-linear evolution of galactic structure. 
A standard assumption is that the course-grained phase-space distribution of the macroscopic simulation particles follows the same evolution equation as that of the fundamental dark matter particles.  This Letter tests this assumption directly for the case of frequent dark matter scatterings, demonstrating that 
this is not generically true.
Specifically, we develop a first-principles map from a microscopic particle physics description of self-interacting dark matter to a representation of macroscopic simulation particles for theories in the short-mean-free-path regime. 
Using this procedure, we show the emergence of an effective force between the simulation particles and derive their interaction cross section, which depends on the one from fundamental particle physics.  
This work provides the first explicit map from particle physics to simulation, which will facilitate exploring the phenomenological implications for galactic dynamics.

\end{abstract}

\section*{$\vphantom{Paper}$} 
\vspace{-16pt}

\noindent \emph{Introduction.}~The intricate structure of the Standard Model of particle physics has inspired many to hypothesize by analogy that what we call ``dark matter'' is actually one or more remnants of a complex ``dark sector.'' 
This generically predicts that the dark matter (DM) will experience non-gravitational self interactions~\citep{Tulin:2017ara}, which could provide an explanation to several small-scale tensions~\citep{2017ARA&A..55..343B, 2022NatAs...6..897S,Adhikari:2022sbh}. 
Studying the imprints of DM self interactions on these scales requires the use of cosmological simulations of galaxy  formation~\citep{2020NatRP...2...42V}.  Due to resolution limits, these simulations track the interactions of DM ``macro-particles,'' which are massive clusters of the microphysical DM states.  This paper provides a first-principles mapping of the fundamental particle physics to its effective implementation in the simulations for the specific example of frequent scatterings, where the DM can be treated as a fluid.  Such a link has never been demonstrated before for any DM model, and it is essential for providing a robust relation between fundamental particle physics parameters and simulation outputs.

DM self scattering enables heat to flow throughout a galactic halo, in stark contrast to the case of purely gravitational interactions~\citep{2000PhRvL..84.3760S}.  This, in turn, affects how the halo's density  distribution evolves with time.  For a halo initialized to a Navarro-Frenk-White profile~\citep{1997ApJ...490..493N}, the heat initially flows inwards, raising the temperature of the central regions and forming an isothermal core~\citep{2000ApJ...543..514K, 2000ApJ...544L..87Y, 2014PhRvL.113b1302K}.  This first stage, referred to as the ``core-expansion'' phase, is not a steady-state solution.  Eventually, the heat flow reverses direction and the core starts to shrink in size and increase in density.  Due to the negative specific heat of the system, the core temperature continues to increase, accelerating the heat flow outwards.  This leads to a runaway process known as ``gravothermal collapse,'' where the core density continues to rise as its radius shrinks~\citep{Balberg:2002ue, 2011MNRAS.415.1125K,2019PhRvL.123l1102E, 2020PhRvD.101f3009N, Yang:2022hkm, 2022MNRAS.513.4845Z, Yang:2022zkd,2023MNRAS.526..758Z, 2024JCAP...02..032Y}. The collapse process can proceed uninhibited until the halo eventually collapses to a black hole~\citep[e.g.,][]{2015ApJ...804..131P}.

Self-interacting dark matter~(SIDM) in the core-expansion phase has been studied in cosmological DM-only simulations~\citep{2012MNRAS.423.3740V, 2013MNRAS.431L..20Z, Rocha:2012jg, 2013MNRAS.430..105P} as well as those with  baryons~\citep{2014MNRAS.444.3684V, 2015MNRAS.452.1468F, 2017MNRAS.472.2945R, 2017MNRAS.469.2845D, 2018MNRAS.476L..20R, 2019MNRAS.490..962F, 2019MNRAS.484.4563D, 2019MNRAS.490.2117R, 2019MNRAS.488.3646R,2021MNRAS.507..720S, 2024A&A...687A.270R}.  Comparatively fewer simulations exist of SIDM halos in the gravothermal collapse regime~\citep{2021MNRAS.505.5327T, 2023ApJ...949...67Y, 2025MNRAS.536.3338C,2025ApJ...986..129N, 2025A&A...697A.213D,  2025ApJ...991...69N}, with progress slowed by numerical challenges in modeling regions of high DM density~\citep{2024JCAP...09..074P,2024PhRvD.110l3024M,2024A&A...689A.300F, 2025A&A...703A.234F}.  In general, all these works are underpinned by the basic assumption that the  interactions of the simulation macro-particles are identical to those of the fundamental DM particles.  Said another way, it is assumed that the coarse-grained phase space representing a simulation macro-particle follows the same evolution as the fine-grained particle phase space~\citep{Rocha:2012jg}.  This is a highly non-trivial assumption, however, as each simulation macro-particle is a conglomerate of individual DM states interacting among themselves.  The primary goal of this work is to scrutinize this assumption.
We develop a novel first-principles formalism that, under a given set of assumptions detailed in this Letter, provides a direct mapping from the particle physics model to the interactions of simulation macro-particles.

Specifically,
we demonstrate that such a formalism exists for the regime where the self interactions dominate the halo evolution.  To define the region of validity of the expansions used here, we introduce the key quantity  
%A key quantity when 
%modeling 
that characterizes the relative strength of
the DM self interactions, namely the Knudsen number: 
\begin{align}
{\text{Kn}}\equiv\frac{\lambda_{\rm mfp}}{\lambda_{\rm J}}\,,
\label{eq:Kn}
\end{align}
which compares the mean-free-path of the DM particles, $\lambda_{\rm mfp}$, with the Jeans length of the system, $\lambda_{\rm J}$.  In the short-mean-free-path~(SMFP) regime, $\text{Kn} \lesssim 1$,  self scatterings are frequent, and the DM can be treated as a fluid. The SMFP regime is typically associated with large scattering cross sections and/or large DM densities.  Depending on the halo and SIDM parameters, some cores can be in the SMFP regime during the expansion phase, while others may only enter this phase during the last stages of gravothermal collapse.

The rest of this Letter is organized as follows. We begin by motivating a particle physics model for DM self interactions that results in SMFP scattering at some point during a halo's evolution.  We then discuss how the DM interactions in this regime can be modeled in a galaxy simulation, starting by dividing the halo's phase space into distinct cells and ultimately course graining those cells to define a simulation macro-particle.  This formalism allows us to model the interactions between the macro-particles:
%using 
we show that this leads to
an effective force and provides a direct relationship between their scattering cross section and the 
one derived from particle physics.
%particle physics one.  
Lastly, we compare to previous results in the literature and conclude. An appendix is provided with the complete derivation of the results presented in this Letter. 

\secspacing

\noindent \emph{The SMFP Regime.} DM self interactions are a generic feature of dark sector models with light mediators, which additionally introduce non-trivial velocity dependence for the scattering rates.   
We take the canonical model where the DM are spin-1/2 
particles with mass $\mdm$ that interact with each other via a Yukawa potential with coupling strength $\sqrt{4 \pi \adm}$ that results from the exchange of a parametrically light spin-1 particle with mass $\mm$~\citep[e.g.,][]{Pospelov:2007mp, 2009JCAP...07..004F, Loeb:2010gj,Tulin:2013teo, 2016PhRvL.116d1302K}.  This yields a long-range force between DM particles.
In the Born limit $(\adm \mdm/\mm \ll {2\pi})$, the self-interacting cross section for distinguishable particles is
\begin{equation}
 \frac{\text{d}\sigma}{\text{d}\cos{\theta}} =\frac{1}{2} \frac{\sigma_0}{\left[1+\frac{v_{\rm rel}^2}{2 w^2} \left(1-\cos\theta\right)\right]^2}\,,
 \label{eq:Bornxs}
\end{equation} 
where $v_{\rm rel}$ is the relative velocity of the DM particles and $\theta$ is their scattering angle. The normalization of the cross section is $\sigma_0 = 4 \pi \adm^2/(\mdm^2 w^4)$ and $w =  \mm/\mdm$ is a characteristic velocity, with $c = 1$.

\begin{figure}
    \centering
    \includegraphics[width=1.\linewidth]{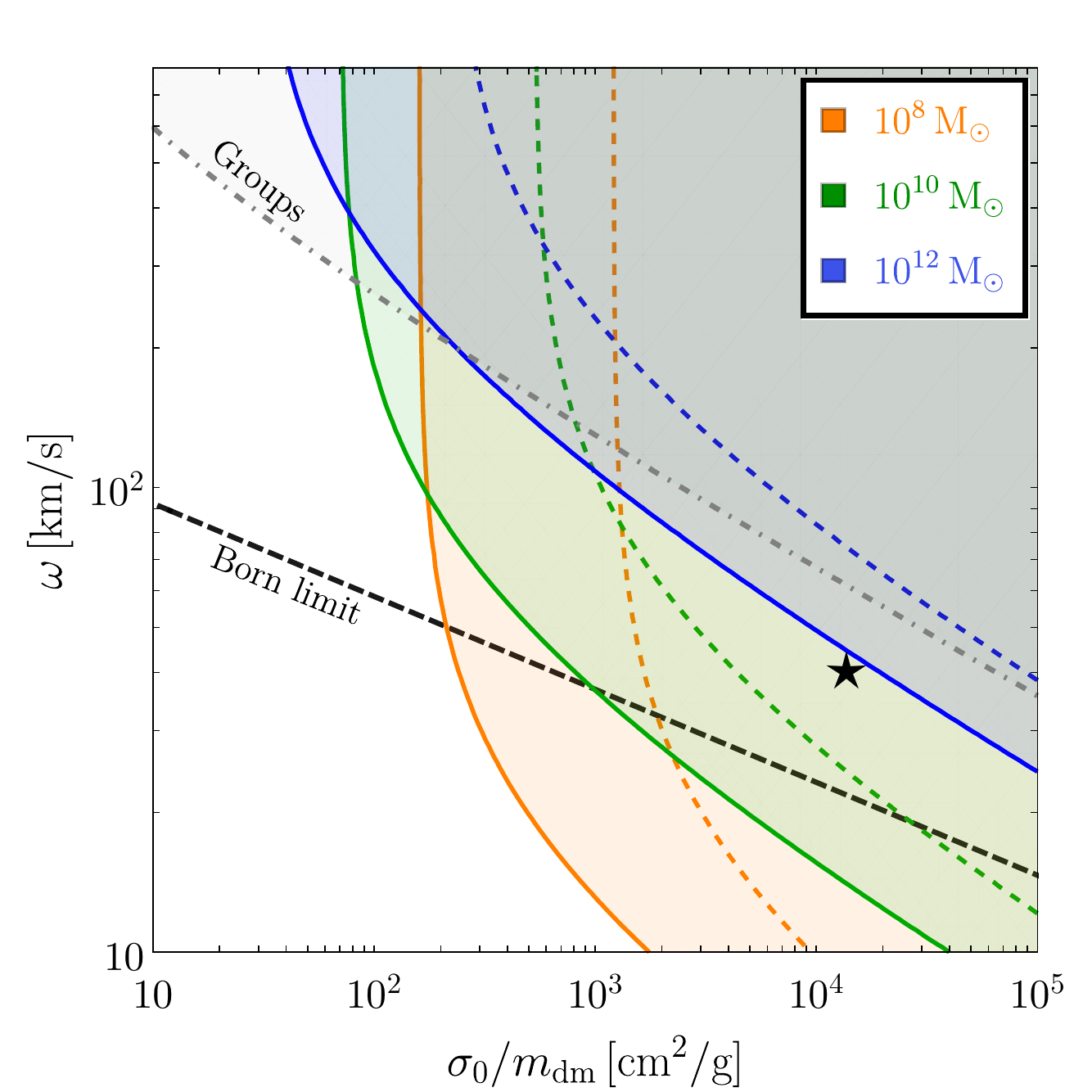}
    \caption{The shaded regions of parameter space to the right of the colored lines denote the SMFP
    regime, where DM scatterings can be implemented as a collective long-range force between simulation macro-particles. The dashed colored lines are evaluated at the stage in which the halo reaches the maximal core size, with $(\rho_c,\, v_c) \simeq (2.4\s \rho_s,\,0.64\s   v_{\rm max})$, while the solid colored lines are defined at a later stage of core collapse, when
    $( \rho_c,\, v_c) \simeq (100\s \rho_s,\,0.75\s  v_{\rm max})$. The orange, green and blue lines refer to virial halo masses of $10^8$, $10^{10}$, and $10^{12}~{\rm M}_{\odot}$, respectively.  In the parameter space above the black dashed line, the Born approximation holds and the Yukawa model can explain the observed DM abundance via the irreducible thermal freeze-out into the mediator. 
    A benchmark point with $\mdm=1\,\text{GeV}$ is represented by a star.  The gray dash-dotted line shows the constraints
    on constant cross sections from galaxy groups from~\cite{Sagunski:2020spe}, re-interpreted for the velocity-dependence predicted by~\cref{eq:Bornxs}.
    }
    \label{fig:smfp}
\end{figure}

In \cref{fig:smfp}, we use the condition $\text{Kn} \sim 1$ to determine the boundary of the SMFP regime (the shaded regions) in the $\omega$ versus $\sigma_{0}/\mdm$ plane. We take different benchmark virial masses and evaluate the Knudsen number at two stages of the halo evolution: at the maximal core stage (dashed colored lines) and 
at a later stage of core collapse
(solid colored lines). Specifically, we consider 
that the halo central density and the 1D dispersion velocity are given by $(\rho_c,\, v_{\rm 1D}) \simeq (2.4\s \rho_s,\,0.64\s  v_{\rm max})$ and $( \rho_c,\, v_{\rm 1D}) \simeq (100\s \rho_s,\,0.75\s  v_{\rm max})$ respectively, following~\cite{Outmezguine:2022bhq}. The corresponding scale density, $ \rho_s$, and maximal rotational velocity, $ v_{\rm max}$, are computed for a NFW profile, using the best-fit mass concentration relation obtained by~\cite{Dutton:2014xda}.

In the same plot, we include the 95\% confidence-level bounds from galaxy groups, obtained by~\cite{Sagunski:2020spe}. Bounds from galaxy clusters are weaker in the relevant parameter space and constraints from dwarf galaxies~(not shown) exist at smaller cross sections than those relevant for our discussion~\citep{Slone:2021nqd, 2023MNRAS.518.2418S}.  The group constraints were obtained assuming velocity- and angular-independent cross sections, which we re-interpret for the transfer cross section (defined in \cref{eq:sigT}), evaluated at $v_{\rm rel} = \left<v_{\rm rel}\right>$ for DM particles following a Maxwell-Boltzmann distribution. 
The inclusion of these bounds assumes that the density profiles of rare and frequent DM self-interaction models are similar.
This assumption should be verified directly, particularly for the parameter space that corresponds to large cross sections relevant for the SMFP regime.

In the same plot, the black dashed line delineates the ``Born limit'' ($\adm/\omega = 2\pi$), below which the approximation used to compute the cross section in \cref{eq:Bornxs} is no longer valid. For the benchmark point marked with a star, the DM candidate explains all of the relic abundance observed for $\mdm=1$\s GeV. This assumes that the abundance is set by the irreducible annihilation of the DM into the mediator, which fixes ${\adm \sim 4\times 10^{-5} \s \mdm/\text{GeV}}$, see~\cite{Feng:2009hw,Tulin:2013teo}. Additional annihilation channels (for example, via visible decays of the mediator) could enhance the cross section, which would decrease the associated relic density.

To summarize, \cref{fig:smfp} demonstrates that there is motivated parameter space where the halo evolution falls within the SMFP regime for the simple illustrative model studied here. This is the region for which it is a good approximation to 
study the DM dynamics
using a fluid approach.
We further note that the SMFP parameter space 
also extends to
more advanced stages of core collapse.

\begin{figure*}[t]
    \centering
    \includegraphics[width=0.9\textwidth]{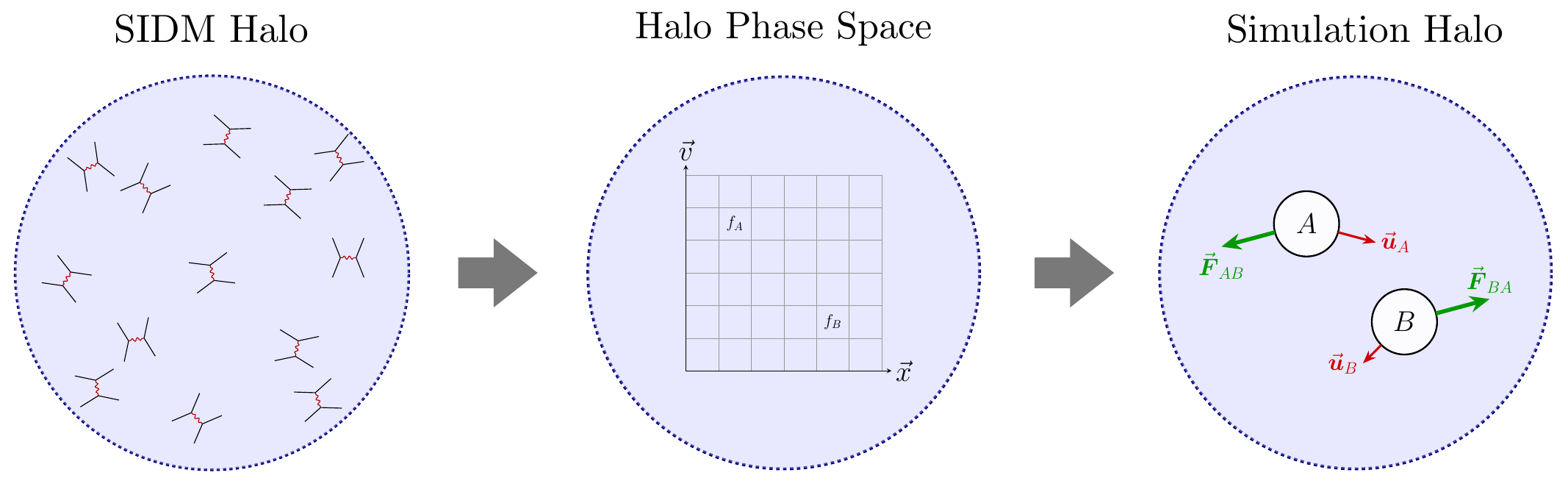}
    \caption{This figure provides an illustration of the coarse-graining procedure introduced in this Letter.  We start with a micro-particle physics model of SIDM.  Assuming we are in the short mean-free-path (SMFP)/frequent collision regime, we can approximate the SIDM as a fluid described by distribution functions (DFs) that satisfy a Boltzmann equation.  We then partition the phase space described by the DFs into cells, which we interpret as the mathematical precursors to the simulation macro-particles.  The final step is to coarse grain these cells, which yields an effective force that describes the impact of the DM self interactions between the simulation macro-particles.  A mathematical summary is given in \cref{eq:scheme} in the appendix.}
    \label{fig:BigPicture}
\end{figure*}

\secspacing

\noindent \emph{Course-Grained Effective Theory.}
For the Yukawa model leading to~\cref{eq:Bornxs}, scatterings in the forward direction are favored as long as $v_{\rm rel}/\omega\gtrsim 1$\s. The DM particles are therefore expected to undergo many small-angle scattering
events that can~\textit{collectively} impact galactic structure.  Here, we provide an effective description that shows how these  scatterings translate to the behavior of the simulation macro-particles, specifically through a collective force acting on each macro-particle.

The appendix of this Letter presents the first-principles derivation of this effective force, starting from a particle physics model. The main steps of the derivation are illustrated in \cref{fig:BigPicture}. We begin by assuming that the SIDM model is in the SMFP regime, and we can describe the system using a Boltzmann equation~(BE). 
%which governs the evolution of distribution functions~(DFs), $f$. 
We treat the galaxy as a collection of phase-space cells (labeled $A,B,\dots$), which we interpret as the mathematical precursors to the simulation macro-particles. The cell labeled by $A$ has a well-defined particle distribution function (DF), $f_A$. Its leading approximation is given by the Maxwell-Boltzmann equilibrium DF, $f_A^{(0)}$. We study the relaxation to equilibrium of $f_A$ due to particle collisions among different cells. To this aim, we perform the ``Chapman-Enskog expansion,'' which computes deviations from equilibrium as an expansion in the Knudsen number, and we additionally assume small particle velocities with respect to the bulk velocity of the halo. We then course grain the particle BE, by means of position and velocity kernel functions. This enables the identification of these cells with macro-particles, with a defined position and velocity. 

In this process, phase-space cell $A$ becomes associated with a simulation macro-particle with velocity $\cg u_A$ and position $\cg x_A$ (and similarly for other cells), as illustrated in~\cref{fig:BigPicture}. Note that we use the bold font to denote coarse-grained (macro-particle) quantities. Explicitly, we obtain the following coarse-grained BE: 
\begin{equation}
    \frac{\partial \cg f_A^{(0)}}{\partial t} + \cg u^i_A \frac{\partial \cg f_A^{(0)}}{\partial \cg x_A^i} = -\sum_{\bar A} \frac{\cg F_{A \bar A }^i}{\mdm} \frac{\partial \cg f_A^{(0)}}{\partial \cg u_A^i} +\mathcal{O}\big(\cg \beta \cg u_A^2\big) \,,
    \label{eq:CGevolution}
\end{equation}
where $\cg f_A^{(0)}\equiv \cg f^{(0)} (\vcg x_A, \vcg u_A,t)$ is the lowest-order term in the coarse-grained DF for macro-particle $A$, $\cg F_{A \bar A }$ is the force between macro-particles $A$ and $\bar{A}$, and we assume Einstein summation convention for the vector indices. In particular, $\cg f_A^{(0)}$ is given by the macro-particle Maxwell-Boltzmann DF, and $\cg \beta$ is a coarse-grained dispersion velocity. For an isotropic halo, ${\left< \cg u_A^2\right> =   3/2\s\cg \beta^{-1} = 3 \cg \s v_\text{1D}^2}$; we 
work in
a frame  where $\left< \vcg u_A\right> = \v{0}$.  

The expansion in \cref{eq:CGevolution} is taken to leading order in the Knudsen number, defined in \cref{eq:Kn} and rewritten here in terms of the macro-particle transfer cross section:
\begin{align}
{\text{Kn}} = \sqrt{\frac{4\pi \s G}{ \rho \s  v_{\rm 1D}^2}}\left(\frac{\cg \sigma_T}{\mdm}\right)^{-1}\,,
\end{align}
where $\rho$ is the halo mass density. This cross section emerges from the computation of the (fundamental) particle collision term at leading order in our expansion and is given by
\begin{equation}
    \frac{\text{d} \cg \sigma_T}{\text{d}\cos{\theta}} = \frac{\int \text{d} v_{\rm rel}^*\s e^{-v_{\rm rel}^{*\,2}/2} v_{\rm rel}^{*\,5} \frac{\text{d}\sigma_T}{\text{d}\cos{\theta}} }{\int \text{d}v_{\rm rel}^*\s e^{-v_{\rm rel}^{*\,2}/2} v_{\rm rel}^{*\,5}}\,,
    \label{eq:sigmaeff}
\end{equation}
where $v_{\rm rel}^* = v_{\rm rel}\s \cg \beta^{1/2}$ is a dimensionless velocity. The normalization is chosen so that a
cross section without any angular and velocity dependence integrates to $\sigma_0$, as defined in \cref{eq:Bornxs}.

The right-hand-side of \cref{eq:CGevolution}
represents the collective force that acts on a given macro-particle $A$ due to collisions with particles \emph{within} macro-particle $B$.  We derive its explicit form to be
\begin{align}
    \cg F_{AB}^i 
    & =\cg {\rho}_B\,\frac{\left(\cg u_B^i - \cg u_A^i\right)}{\cg \beta^{1/2}}  \frac{8\sqrt{2\pi}}{3}   \int \text{d}\!\s\cos{\theta}\frac{\text{d} \cg \sigma_T}{\text{d}\cos{\theta}}\,,
    \label{eq:Feff}
\end{align}
where $\cg \rho_B$ is the macro-particle mass density.  We underscore that the force on the simulation macro-particles properly accounts for the particle DM collisions among different  phase-space cells. Additionally, because the macro-particle DF satisfies a  BE (see \cref{eq:CGevolution}), we can directly read off the velocity of each macro-particle.

The assumptions that underlie the derivation of our central result in~\cref{eq:Feff} are
\begin{enumerate}
\item Collisions bring the system near equilibrium. This allows us to write the particle DF as an expansion near the Maxwell-Boltzmann distribution, $f = f^{(0)} + {\rm Kn}^k f^{(k)}$. We then use the Chapman-Enskog expansion to find the first approximation to equilibrium, $f^{(1)}$, and explicitly write the collision term to this order; see App.~\ref{App:A}.
\item Macro-particles have small velocities near equilibrium. The effective force obtained in~\cref{eq:Feff} is the linear velocity response of the system to a temperature gradient.
\item Macro-particles have the same mass and they occupy the
same volume of phase space before and after the collision, so $\cg \rho_A = \cg \rho_B$\,. 
\item Macro-particles are localized in space. This means we assume a trivial $\delta$-kernel to coarse grain in position space.   This  trivially maps ${\rho_A},\s\beta$ to their coarse-grained values, ${\cg \rho_A},\s\cg \beta$. 
\item Macro-particle momentum is conserved in the collision. As shown in App.~\ref{app:B}, this assumption points to a particular form of the kernel in velocity space, which we also assume to be spherically symmetric. An example of such a kernel is a Gaussian times a linear function of velocity, see \cref{eq:Wg}. We verified that this kernel also trivially maps the gravitational force between the particle and macro-particle descriptions, see \cref{eq:CGgravity}.

\end{enumerate}

\begin{figure}[t!]
    \centering
    \includegraphics[width=1.\linewidth]{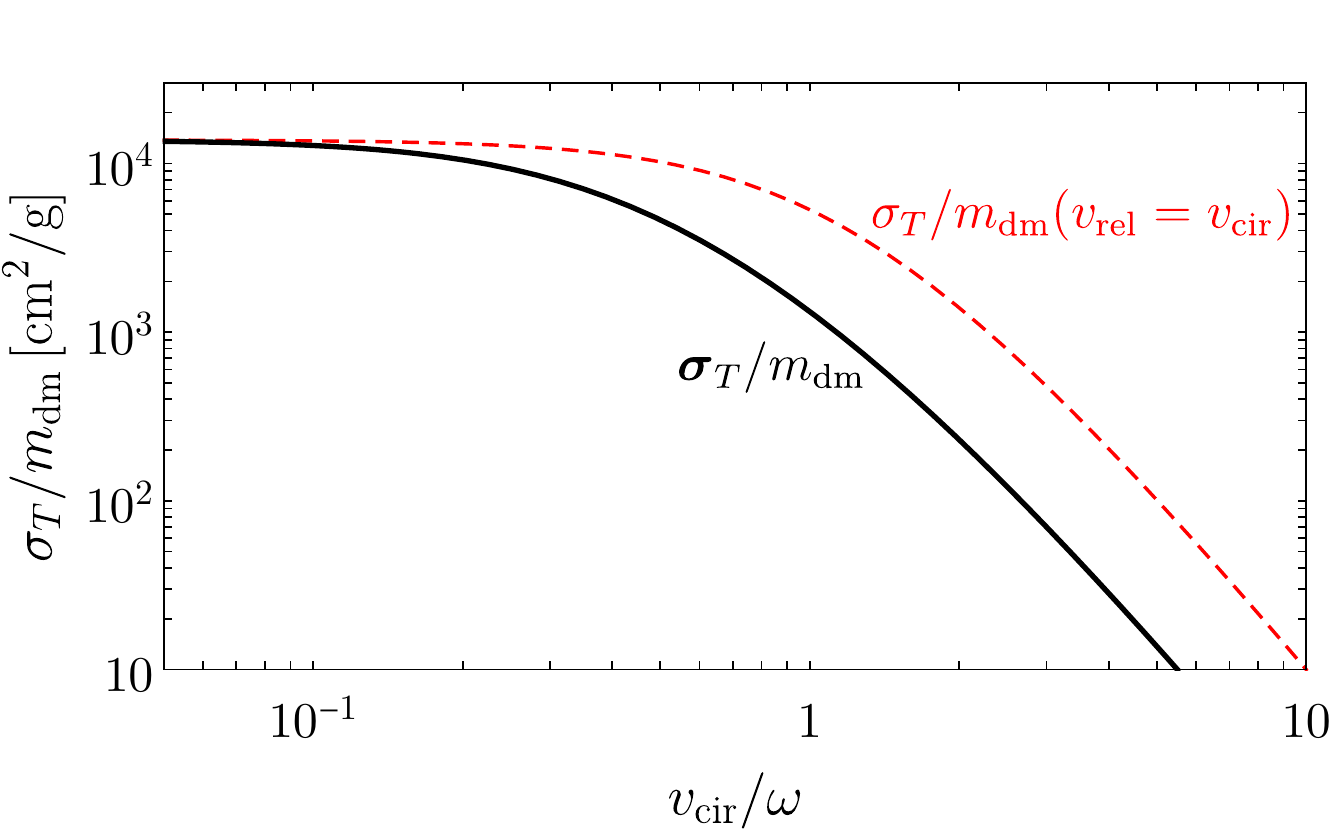}
    \caption{This shows the dependence of the effective macro-particle [black solid line] and micro-particle [red dashed line] cross sections on $v_{\rm cir} \sim \sqrt{G M(r)/r} \sim \sqrt{2} v_{\rm 1D}$ for an isothermal halo profile. 
    We set $\sigma_{0}/\mdm = 1.4\times 10^4\,\text{cm}^2/\text{g}$ and $\omega = 40\,\text{km}/\text{s}$. This benchmark model is represented by a star in~\cref{fig:smfp}.
}
    \label{fig:sigT}
\end{figure}

To appreciate the physical implications of course graining, we can compare the effective cross section from~\cref{eq:sigmaeff} against the naive cross section that one would obtain without implementing a coarse-graining scheme. To evaluate the naive cross section, we need to make a choice for how to evaluate the velocity. We assume $\g = v_{\rm cir} \equiv \sqrt{G M(r)/ r}$, following e.g.,~\cite{Harvey:2025paa}). For an isothermal halo, which we consider here for simplicity, $v_{\rm cir} = \sqrt{2} \s v_{\rm 1D}$. 
\cref{fig:sigT} compares our effective and the naive cross sections as a function of $ v_{\rm cir}$, for the DM model represented by a star in~\cref{fig:smfp}. For virial masses around $10^{10}\,\rm M_\odot$, with typical velocities of order $ v_{\rm cir}\sim 40$\,km/s, the naive approach overestimates the cross section by a factor $\sim 10$. As expected, when $w\gg v_{\rm cir}$ the two cross sections become equal since $\sigma\to \sigma_0$ in this limit.

\secspacing
\noindent \emph{Comparison to Previous Work.} 
Prior studies have used a semi-analytical approach based on the gravothermal fluid formalism to investigate SIDM halo evolution in the SMFP regime---see e.g.~\cite{Yang:2022hkm,Outmezguine:2022bhq,Gad-Nasr:2023gvf}. We now comment on the comparison between our work and their approach.
Both approaches assume a fluid description of the halo and identify that core formation/collapse results from the existence of a temperature gradient.  Our formalism maintains a notion of temperature because we work near equilibrium: temperature is determined by the velocity dispersion. 

That being said, our approach is fundamentally different from the gravothermal fluid formalism. The latter treats the halo as a mono-component gas, whereas we treat it as a multi-species gas, where the species correspond to the phase-space cells that are the intermediate step towards defining a representation of the simulation macro-particles.  This intermediate step is necessary to capture the behavior of the macro-particles, which is not the goal of the gravothermal fluid formalism. This key difference yields fundamentally different results for the cross section mapping.

At the level of thermodynamics, these two descriptions are also different. For a mono-component gas, the leading thermodynamic flux generated by a temperature gradient is the heat flux, i.e.,~the third moment of the DF, but this is not true here. In our approach, a temperature gradient leads to both diffusion and heat flux, the first and third moment of the DF, now computed with respect to macro-particle velocities (instead of the bulk velocity of the galaxy, which we neglect). Therefore, when treating macro-particles, this diffusion flux is the leading thermodynamic effect assuming a small expansion in velocity such that higher-order moments of the DF can be neglected.

It is this diffusion flux that maps onto an effective force for the macro-particles. 
Additional effects, such as the heat flux---which one can interpret as the random heat conduction within the macro-particle---are higher-order in this expansion. We leave understanding these higher-order effects, and the extent to which they can be accommodated by modifying the effective force introduced here, to future work.

There is another related line of research in the literature. \cite{Kahlhoefer:2013dca} present a heuristic effective drag force that is estimated by evaluating evaporation rates for individual DM particles in a halo.  This force has been implemented in simulations by~\cite{Fischer:2020uxh, Sabarish:2023ija,Arido:2024hfk} 
to model anisotropic scatterings with large cross sections. These works assume that the drag force derived at the particle-level generalizes to the simulation macro-particles.  Our work, in contrast, presents a systematic expansion of the BE to derive an effective force that applies in the SMFP regime for simulation macro-particles. The collective force derived from first principles gives a universal mapping between a velocity-dependent particle cross section and a constant macro-particle cross section, which depends linearly on the macro-particle velocity. This behavior follows from the fact that in the SMFP regime, a given particle in macro-particle $A$ experiences a collective effect due to all possible scatterings with particles inside $B$.  This is one of the key conclusions of this Letter.  

\secspacing
\noindent \emph{Conclusions.}  Galaxy simulations serve as an important tool for studying the implications of different DM models on small scales.  This Letter tests a standard assumption made in the literature:~that the phase-space distribution of the simulation macro-particles follows the same evolution equation as that of the fundamental DM particles in the halo.  We show that this assumption is \emph{not} valid, at least  for the case of self-interacting DM models in the SMFP regime.  The inconsistency arises because, when two macro-particles interact, all of their constituent DM states do as well, and this effect must be properly accounted for.  We demonstrate for the first time that a first-principles formalism does exist to map the particle-level interactions to the macro-particle interactions in the SMFP regime. The results of this work motivate careful study extending the formalism to other DM scenarios, as well as implementation in numerical simulations.  Such work is critical for drawing robust conclusions regarding specific particle physics models using galactic-scale observables.

\begin{acknowledgments}
We thank Andrea Caputo, Guilherme Guedes, Felix Kahlhoefer, Ben Safdi, Kai Schmidt-Hoberg, Oren Slone, and Haibo Yu for valuable discussions.
T.C.\ is supported by the US Department of Energy under grant DE-SC0011640.
M.L.\ is supported by the US Department of Energy under
grant DE-SC0007968 and by a Simons Investigator Award. 
\end{acknowledgments}

\appendixTC

\appendix

\noindent This appendix provides the detailed derivation of the claims made in the main text.
We treat the galaxy (with bulk velocity $\vec U$) as a collection of phase-space cells, which we interpret as the mathematical precursors to the coarse-grained simulation macro-particles. The Boltzmann equation~(BE) governing the evolution of these cells is analogous to that of a multi-species gas, which has a more general thermodynamic description than a mono-component gas. We focus on the physics that governs the relaxation to equilibrium of the particle distribution function~(DF) from collisions between different cells, using the ``Chapman-Enskog expansion.'' We then course grain the BE for these cells by introducing kernel functions in position and velocity space. This enables the identification of such cells with macro-particles, which have a corresponding velocity and position, and are associated with an effective force. This effective force provides a way to model the impact of dark matter~(DM) self interactions in the short-mean-free-path~(SMFP) regime within a simulation framework.

In slightly more detail, the key steps are as follows:  
\begin{align}
\includegraphics[valign=c, width=40pt]{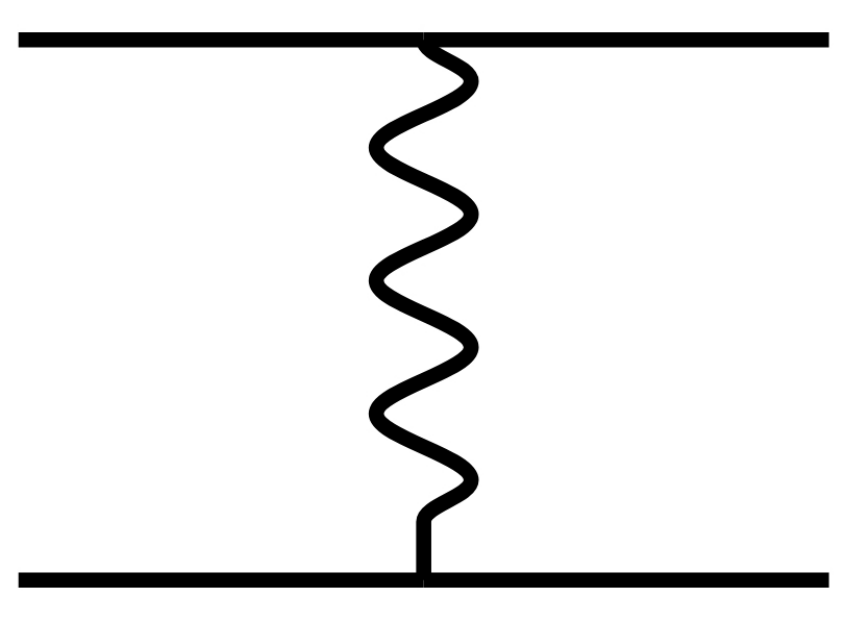}
\quad\xrightarrow[\text{\cref{eq:BEdimless}}]{\,\,\rm BE\,\,}\quad   {\frac{\text{d} f}{\text{d} t}} = C[f] \quad\xrightarrow[\text{\cref{eq:firstapprox}}]{\,\,\rm SMFP\,\,} \quad {\frac{\text{d} f^{(0)}}{\text{d} t}} = \mathcal{O}(\text{Kn}^0)
\quad\xrightarrow[\text{\cref{eq:CGevolutionAppendix}}]{\,\,\rm CG\,\,}  \quad
{\frac{\partial \cg f}{\partial t}} + \cg u^i  {\frac{\partial \cg f}{\partial \cg x^i}} = - {\frac{\cg F^i}{\mdm}}  {\frac{\partial \cg f}{\partial \cg u^i}} \,,
\tag{A}
\label{eq:scheme}
\end{align}
where bold text refers to course-grained quantities while arrows refer to vector quantities throughout.
Starting with a particle physics Lagrangian for a DM particle of mass $m_{\rm dm}$, we compute the cross section which is an input to the BE that describes the evolution of the particle DF, $f$, in terms of the collision integral, ${\rm C}[f]$. In the SMFP regime, where the Knudsen number $\text{Kn}< 1$ (see \cref{eq:Kn}), we perform the Chapman-Enskog expansion about the equilibrium DF.  This relates the collision integral at first order in $\text{Kn}$ with the Maxwell-Boltzmann DF, $f^{(0)}$. The final step is to coarse grain the leading-order result to derive a BE for the macro-particle DF, $\cg f$. This requires introducing coarse-graining kernel functions in position and velocity space. The final equation has the interpretation that particle collisions can be recast as an effective force $\vec{\cg F}$ at leading order in the velocity expansion.  Each macro-particle is associated with position $\cg x^i$ and velocity $\cg u^i$, where the $i$ index refers to the vector components. This force explicitly satisfies macro-particle momentum conservation. The rest of the appendix is devoted to showing this derivation in detail. From here on, we use Einstein summation convention for spacetime indices and we explicitly write summations over phase-space cells.

\vspace{0.5cm}

\section{Chapman-Enskog expansion on a Multi-cell phase space}\label{App:A}

\noindent
We start from the distribution of fundamental DM particles in the galactic halo.  In a state of equilibrium, the particle DF follows a specific velocity distribution. If we define $\mathcal N$ phase-space cells (labeled as %`species'
$A,B,\dots$) in such a state, as long as the number of particles/cell is sufficiently large, the velocities internal to the cells  %their velocities 
will follow the same distribution. Otherwise, the global state of the system would not be equilibrium. One then defines a particle DF for each cell $f_A (\vec x_A,\vec v_A, t)$, which follows its own BE (in dimensionless form):
\begin{equation}
    \frac{\partial f^*_A}{\partial t^*} + \vec v^{\s *}_A \frac{\partial f^*_A}{\partial \vec x^{\s *}_A} - \frac{\Phi_c}{v_c^2} \vec \nabla^* \Phi^* \frac{\partial f_A^*}{\partial \vec v_A^{\s *}}  = \frac{x_c}{\lambda_{\rm mfp}}\sum_{\bar A = A,B,\dots}\int \text{d} \Omega \int \text{d}^3  v_{\bar A}^* \s \frac{\text{d}\sigma^*}{\text d \Omega} \big[f_{A}^{\prime *} f_{\bar A}^{\prime *}-f_A^* f_{\bar A}^*\big]\,\big|\vec v^{\s *}_{\bar A} - \vec v^{\s *}_{A}\big|\, \equiv C\big[f_A^*\big]\,,
    \label{eq:BEdimless}
\end{equation}
where $\Phi^*$ is the dimensionless gravitational potential, $\Omega$ is the solid angle for scattering, $\text{d}\sigma^*/\text{d}\Omega$ is the dimensionless differential scattering cross section for self-interacting DM, and $x_c,\,v_c$, and $\Phi_c$ are characteristic length, velocity, and potential scales (to be defined later).  Here, the particle coordinates, $\v x_A$ and $\vec v_A$, fall within a given cell $A$, whose limits will be defined by the coarse-graining scheme. The sum on the right-hand-side of this equation takes into account both intra- and inter-cell collisions. We use the notation $f_{A}^{(\prime)} \equiv f_{ A}^{(\prime)} \big(\vec x_{ A}^{\vphantom{(\prime)}},\vec v_{ A}^{\s(\prime)},t\big)$, where primes are used 
for post-scattered velocities. The $f_{ A}^{(\prime)}$ satisfy the usual momentum and energy conservation relations. The length scale that arises is the particle mean free path (MFP):
\begin{align}
\lambda_{\rm mfp} \equiv \frac{1}{n\s {\sigma_c}}\,, 
\end{align}
which depends on position via the total number density $n$ and on a self-interaction cross section $\sigma_c$, which will be obtained explicitly by evaluating the collision term. Below, we identify $\sigma_c$ with the effective cross section given in~\cref{eq:xs}. The dimensionless variables are normalized with respect to characteristic scales: $x^* \equiv x/x_c$, $v^* \equiv  v/v_c$, $\Phi^* = \Phi/\Phi_c$, $\sigma^* \equiv \sigma/\sigma_c$, and $f_A^* \equiv f_A/(f_A)_c $ with $(f_A)_c = n_A/v_c^3$. This is true since~$\int  \text{d}^3 x_A \s \text{d}^3 v_A \s f_A = N_A$, the number of particles in a cell, which implies $x_c^3 \s v_c^3 \s (f_A)_c = x_c^3 \s n_A $, where $n_A$ is the number of density in a cell (normalized to the same volume as the total number density). Therefore, an overall $n_A$ comes out of the dimensionless collision integral, which becomes $n$ after considering the sum over all cells. This explains the definition of the MFP.

We identify $v_c\sim v_{\rm 1D}$, which is the one-dimensional velocity dispersion of the DM particles in the halo. We assume the same dispersion velocity for all cells in this region. Therefore, $\Phi_c \sim v_c^2$ for a virialized halo. The characteristic (macroscopic) length scale $x_c$ can be derived using the Poisson equation:
\begin{align}
\nabla^2 \Phi = 4\pi\s G \rho \,\, \Rightarrow \,\, x_c \sim \sqrt{\frac{v_c^2}{4\pi G\rho}} \equiv \lambda_{\rm J}\,,
\end{align}
where $\lambda_{\rm J}$ is the Jeans scale, defined in terms of the gravitational constant, $G$, and the halo mass density, $\rho$.

The Jeans scale also controls the variation of the thermodynamic fields $\rho$ and $v$, as can be seen from the
equation of hydrostatic equilibrium~(\cite{Balberg:2002ue}), which we  rewrite as
\begin{equation}
    \frac{1}{\lambda_\rho} \frac{\partial \log{\rho}}{\partial x^*} + \frac{1}{\lambda_{v}} \frac{\partial \log{v_{\rm 1D}^2}}{\partial x^*} =\frac{1}{\lambda_{\rm J}} \frac{\partial \Phi^*}{\partial x^*}\,,
\end{equation}
after defining the pressure $p=\rho\s v^2$ and $\lambda_\alpha$ as the typical variation scale of a given macroscopic field $\alpha$. This equation shows that the temperature, 
$k_B \s T = \mdm \s v_{\rm 1D}^2,$
is expected to vary on scales of the order of $\lambda_{\rm J}$ 
(here $k_B$ is the Boltzmann constant).  
Therefore, at scales much smaller than the Jeans length, thermodynamic gradients are small and all particles share the same temperature, as they must, since they are in local equilibrium.

We focus on the regime where the Knudsen number (Kn) is small:
\begin{align}
{\rm Kn}\equiv \frac{\lambda_{\rm mfp}}{\lambda_{\rm J}} \ll 1\,.
\end{align}
This condition defines the SMFP regime, and we will assume this limit from here forward. Moreover, in this limit, gravitational interactions can be neglected with respect to DM collisions, which means that the third term on the left-hand-side of \cref{eq:BEdimless} can be ignored.

Next, we perturb the system away from equilibrium and compute the leading correction to the particle-level DF in a given cell. We also work with quantities which are now dimension-full. Specifically, we take
\begin{equation}
    f_A^{\vphantom{(i)}} %\big(t,\vec x,\vec v\s\big) 
    = \sum_{k=0}^\infty \s {\rm Kn}^{ k} \s f_A^{({k})}\,,
    \label{eq:CEexp}
\end{equation}
which is known as the Chapman-Enskog expansion.

At $\mathcal{O}({\rm Kn}^{-1})$, the collision term is superleading in the expansion in~\cref{eq:CEexp}, and so we have:
\begin{equation}
C\big[f_A^{(0)}\big] = 0\,,
\end{equation}
which is the condition of detailed balance. This is simply the statement that $f_A^{(0)}$ is the equilibrium DF, which is the Maxwell-Boltzmann distribution:
\begin{equation}
\label{eq:MB2}
    f^{(0)}_{A} (\vec x_A, \vec v_A, t)= n_{A}(\vec x_A, t) \left(\frac{\beta}{\pi}\right)^{3/2} 
    e^{-\beta \left(\vec v_{A} - \vec U\right)^2}\,.
\end{equation}
From here forward, we work with $\vec U = \v 0$ and assume that the halo is isotropic and spherically symmetric. This implies
\begin{align}
\frac{1}{\beta} = 2\s v_{\rm 1D}^2\,.
\end{align}

Next, we analyze the $\mathcal{O}({\rm Kn})$ contribution to the collision term.  It is useful to write the BE in terms of a quantity $\phi^{(1)}$, which is defined by
\begin{align}
f_A^{(1)} \equiv \phi_A^{(1)} f_A^{(0)}\,.
\label{eq:f1}
\end{align}
The $\mathcal{O}({\rm Kn}^0)$ BE is then
\begin{equation}
\frac{\text{d}f_A^{(0)}}{\text{d}t} =
\frac{\partial f_A^{(0)}}{\partial t} + v^i_A \frac{\partial f_A^{(0)}}{\partial x_A^ i}  = \sum_{\bar A = A,B,C,\dots} \int \text{d} \Omega \int \text{d}^3  v_{\bar A} \,\frac{\text{d}\sigma}{\text{d} \Omega}  \left(\phi_{\bar A}^{\prime(1)} - \phi_{\bar A}^{(1)} + \phi_A^{\prime(1)} - \phi_A^{(1)}\right) f_A^{(0)} f_{\bar A}^{(0)}|\vec v_{\bar A} - \vec v_A| \,,
\label{eq:firstapprox}
\end{equation}
where we use $f_A^{\prime \s(0)} = f_A^{(0)}$ due to particle energy conservation. After using the balance equations for mass, momentum, and energy density~(\cite{Kremer2010}), while also only keeping the temperature gradient,~\cref{eq:firstapprox} can be simplified to
\begin{equation}
    \frac{1}{\beta} \frac{\partial \beta}{\partial x_A^i} f_A^{(0)} v_A^i\left(\beta v_A^2 -\frac{5}{2}\right) = \sum_{\bar A = A,B,C,\dots} \int \text{d} \Omega \int \text{d}^3  v_{\bar A} \,\frac{\text{d}\sigma}{\text{d} \Omega} \left(\phi_{\bar A}^{\prime(1)} - \phi_{\bar A}^{(1)} + \phi_A^{\prime(1)} - \phi_A^{(1)}\right) f_A^{(0)} f_{\bar A}^{(0)}|\vec v_{\bar A} - \vec v_A| \,.
\end{equation}
Given that the $\vec{v}_{\bar{A}}$ are integrated over on the right-hand-side of the equation, we can express the solution as the following power series in $\vec{v}_{A}$
\begin{equation}
    \phi_A^{(1)} = \mathcal P_A v_A^i \frac{\partial \beta}{\beta \partial x_A^i} +\mathcal{O}(v_A^2)\,,
    \label{eq:PA}
\end{equation}
where $\mathcal P_A$ is a constant. Here, we focus on the linear response of the system to out-of-equilibrium dynamics, so we neglect higher-order terms in velocity in the expression above. By combining \cref{eq:f1} and \cref{eq:PA}, we can match $\mathcal P_A$ to
the average velocity of a phase-space cell: 
\begin{align}
u_A^i = \frac{1}{n_A}\int \text{d}^3 v_A \s v_A^i f^{\vphantom{i}}_A\,,\qquad \text{with} \qquad n_A = \int \diff^3 v_A\,  f_A\,,
\label{eq:averages}
\end{align}
where $n_A = \rho_A/\mdm$ is normalized with respect to the entire spacial volume. One can show that $n_A$ is conserved in the presence of particle collisions at all orders in Kn~(\cite{Kremer2010}). 
This, together with the small-velocity expansion, justifies the description of phase space in terms of distinct cells, even for an out-of-equilibrium state.

The $\mathcal{O}({\rm Kn}^0)$ linear response captures the effect of the diffusion flux in the $i^{\rm th}$ direction, which is
\begin{equation}
    J^i_A \equiv  \rho_A \s  u_A^i 
    =  \mathcal P_A  \frac{\partial \beta}{\beta \partial x_A^i} \frac{\rho_A}{2\beta} \,,
\end{equation}
and so we have
\begin{equation}
\phi_A^{(1)} = 2\s\beta\s \v u_A \cdot \vec v_A = \frac{2\s \beta}{\rho_A} \s \v J_A \cdot \vec v_A \,.
   \label{eq:DFlinear}
\end{equation}
Note that because we can identify $\vec J_A$ with the average velocity at this order in the expansion, we can proceed without having to evaluate the unknown $\mathcal P_A$ function.

In general, the inclusion of more powers of velocity in $\phi_A^{(1)}$ leads to a contribution to the BE from additional thermodynamic fluxes. This is a well-known aspect of kinetic theory (see e.g.~\cite{Grad1958,Kremer2010}).  For reference, we give the more general form of the DF:
\begin{equation}
\label{eq:Grad}
    f_A = n_A \left(\frac{\beta}{\pi}\right)^{3/2} e^{-\beta v_A^2} \bigg[ 1 + 2\s \beta \frac{J_A^i v_A^i}{\rho_A} + \frac{2\s\beta^2}{\rho_A} \left\{p_{\left<ij\right>}^A v^i_A v_A^j + \frac{4}{5} q^i_A v^i_A \left(\beta\s v_A^2 -\frac{5}{2}\right)\right\}+\dots\bigg]\,,
\end{equation}
where $p$ is the pressure and $q$ the heat flux. The angular brackets denote the trace-less symmetric part of a tensor. The dots represent terms sourced by higher-order moments of the particle DF. In particular, the heat flux is also generated by a temperature gradient. 
However, since we are working to linear order in velocity, we can consistently set the flux to zero.  In other words, we are truncating the expansion following
    \begin{align}
q_A^{i} & = \int \text{d}^3 v_A \frac{m}{2} \big(v_A^j - u_A^j\big)^2 \big(v_A^i - u_A^i\big) f_A =\mathcal{O}\big(\beta \s u_A^2\big)\,. %\mathcal{O}(\beta \s \big(\vec u_A - \v u_{\rm gal})^2\big)\,,
\end{align}

\vspace{0.5cm}
\section{Coarse Graining The Boltzmann Equation}
\label{app:B}
\noindent
Given the $\mathcal{O}({\rm Kn})$ particle DF obtained in~\cref{eq:DFlinear}, we can proceed to coarse grain the collision integral to provide a map onto a model for the simulation macro-particles.  We will show that, at this order in the expansion, the coarse-grained collision integral can be expressed as an effective force.  Specifically, we will justify \cref{eq:CGevolution} in the main text, which is the claim that the macro-particles obey the following leading-order BE:
\begin{equation}
    \frac{\partial \cg f_A^{(0)}}{\partial t} + \cg u^i_A \frac{\partial \cg f_A^{(0)}}{\partial \cg x_A^i} =-\sum_{\bar A} \frac{\cg F_{A \bar A }}{\mdm} \frac{\partial \cg f_A^{(0)}}{\partial \cg u_A^i} \,,
    \label{eq:CGevolutionAppendix}
\end{equation}
 where $\cg f_{A} \equiv \cg f (\vcg x_A,\vcg u_{ A},t)$\,, and bold is used to denote macro-particle variables as above. In particular, $\cg u_A^i$ is the macro-particle velocity, which is the coarse-grained version of the average cell velocity in \cref{eq:averages}. We prove this in the next steps. 
This equation is obtained by averaging the leading-order BE in \cref{eq:firstapprox} against position and velocity kernels, $G$ and $W$ respectively.  The purpose of introducing these kernels is to coarse grain the coordinates $\{\v x_A , \s \v {v}_A \}\to \{ \vcg x_A, \s \vcg u_A \}$, 
thereby mapping the particle DFs to macro-particle DFs, $f_A \to {\cg f}_A$.

Explicitly, the coarse-grained BE at $\mathcal{O}(\text{Kn}^0)$ is 
\begin{align}
 &\mathcal N \int\text{d}^3  x_A\s  G_A (\v x_A, \vcg {x}_A)  \int \text{d}^3 v_A  \s  W_A (\v v_A, \vcg u_A) \left[\frac{\partial f_A^{(0)}}{\partial t} + v_A^i \frac{\partial f_A^{(0)}}{\partial x_A^i}\right]  = \sum_{\bar A = B,C,\dots} \mathcal{I}_{A \bar{A}}\,,
  \label{eq:CG}
\end{align}
with
\begin{align}
\mathcal{I}_{A \bar{A}} &=  \mathcal N \int \text{d}^3  x_A \s G_A (\v x_A, \vcg {x}_A)  \int \text{d}^3 v_A  \s   W_A (\v v_A, \vcg u_A) \int \text{d}^3 x_{\bar A} \s G_{\bar A} (\v x_{\bar A}, \vcg { x}_{\bar A})  \nonumber \\[4pt]
 &\hspace{55pt} \times \int  \text{d} \Omega \int \text{d}^3  v_{\bar A} \,\frac{\text{d}\sigma}{\text{d} \Omega}  \left(\phi_{\bar A}^{\prime(1)} - \phi_{\bar A}^{(1)} + \phi_A^{\prime(1)} - \phi_A^{(1)}\right) f_A^{(0)} f_{\bar A}^{(0)} |\vec v_{\bar A} - \vec v_A| \,.
\end{align}
Here, the total number of cells/macro-particles $\mathcal N$ is introduced for convenience.  This factor will be included in the definition of the coarse-grained DF, such that the total number density is given by
\begin{align}
   n (\vcg x,\s t) = 
   \sum_A \cg n_A(\vcg x,\s t) = \mathcal N \s \cg n_A(\vcg x,\s t) 
   & = \mathcal N \int \text{d}^3  x_{A} G_{ A} (\v x_{ A}, \vcg { x}_A) n_A(\v x_{A}, t)\notag\\[2pt]
   & =\mathcal N \int \text{d}^3  x_{A} G_{ A} (\v x_{ A}, \vcg { x}_A)\int \text{d}^3 v_{A} f_{A} (\v x_{A}, \v v_{ A}, t)\notag\\[2pt]
   & =\mathcal N \int \text{d}^3 \cg u_{A}\s W_{A} (\v v_A, \vcg u_{ A})\int \text{d}^3  x_{A} G_{ A} (\v x_{ A}, \vcg { x}_A)\int \text{d}^3 v_{A} f_{A} (\v x_{A}, \v v_{ A}, t)\notag\\[2pt]
   & =  \int \text{d}^3 \cg u_{A}\s \cg f_A (\vcg x_A, \vcg u_{A}, t) \,,
    \label{eq:nMicroToMacro}
\end{align}
where the total and the macro-particle number densities are normalized to the same volume. Moreover, the kernel functions in \cref{eq:nMicroToMacro} are assumed normalized, 
namely
\begin{equation}
    \int \text{d}^3 \cg u_A \s W_A (\v v_A, \vcg u_A) = 1\,.
\end{equation}
\cref{eq:nMicroToMacro} once again proves that $\cg f$ has the interpretation of a coarse-grained phase-space DF.

We will now show that~\cref{eq:CG} is equivalent at leading order to \cref{eq:CGevolutionAppendix}, with the following normalized kernel in velocity space:
\begin{equation}
    W_A (\v v_A,\s \vcg u_A) = \left(\frac{\beta}{\pi}\right)^{3/2} \big(1+2\s \beta\s  \v v_A \cdot \vcg u_A\big)\, e^{-\beta \cg u_A^2}\,,
    \label{eq:Wg}
\end{equation}
while we take
\begin{align}
G_A  (\vcg x_A,\s  \vec{x}_A) =\delta^3 (\v x_A - \vcg x_A)
\label{eq:GAdelta}
\end{align}
to coarse grain in position space. This choice of position-space kernel has the physical interpretation that the simulation macro-particles have zero size.  In practice, the macro-particles are typically given a non-zero ``softening length.''  This will not impact the conclusions drawn here as long as the integral in position space does not affect the mapping of the particle BE onto \cref{eq:CGevolutionAppendix}.  We leave implementing a position-space kernel that includes a softening length for future work.

First, we show that evaluating the integrals on the left-hand side of \cref{eq:CG} simply coarse grains the collisionless BE. The time derivative term is trivial, so we focus on the velocity-dependent term:
\begin{align}
    \mathcal N \int \diff^3 x_A G_A (\v x_A, \vcg { x}_A)  \int \diff^3  v_A    W_A (\v v_A, \vcg u_A) v_A^i \partial^i f_A^{(0)} & = \mathcal N \s\partial^{i} \cg n^{\vphantom{j}}_A (\vcg {x}_A,t)\s\cg u^j_A \left(\frac{\cg \beta}{\pi}\right)^{3} e^{- \cg\beta \cg u_A^2} \int \diff^3 v_A  \s  \,\big(2\s\cg \beta \s v_A^i v_A^j\big)  e^{- \cg \beta v_A^2} \nonumber \\[3pt]
    & = \mathcal N \s  \partial^{i} \cg n_A (\vcg {x}_A,t)\s  \cg u^i_A \s \left(\frac{\cg \beta}{\pi}\right)^{3/2} e^{- \cg\beta \cg u_A^2} \nonumber \\
    & = \cg u^i_A \partial^i \cg f^{(0)}_A \,,
    \label{eq:macroBE}
\end{align}
where we used \cref{eq:MB2} for $f^{(0)}_A$ evaluated at $U = 0$, and assumed spherical symmetry in velocity space.  In general, the velocity dispersion is position-dependent, so $\beta(\vec{x}) \rightarrow \cg \beta(\vec{\cg x})$ after applying the position kernel. In the equation above, the appearance of the coarse-grained cell velocity is not trivial, and it is what allows the interpretation of $\cg u_A^i$ as the macro-particle velocity, at least for the choice of kernel in~\cref{eq:Wg}.

We now develop the collision term to derive the effective force. Focusing on the collisions between two macro-particles $A$ and $B$, we have
\begin{align}
   \cg{\mathcal I}_{AB} &=  \int \diff \Omega\int \diff^3 v_A \int \diff^3 v_B\bigg[\mathcal N\s \left(\frac{\cg \beta}{\pi}\right)^{3/2} e^{-\cg \beta \cg u_A^2} 2\s\cg\s \beta \s \v v_A\cdot \vcg u_A\bigg]  2\s\cg \beta\, (\cg u_A - \cg u_B)^i (v_A^\prime-v^{\vphantom{\prime}}_A)^i \nonumber \\[3pt]
    &\hspace{115pt} \times \cg n_A\s \cg  n_B  \left(\frac{\cg \beta}{\pi}\right)^{3} e^{-\cg \beta \left(v_A^2 + v_B^2\right)} |\vec v_B - \vec v_A| \frac{\diff\sigma}{\diff\Omega}\,,
\end{align}
after coarse graining the position space. This integral can be simplified to
\begin{align}
   \cg {\mathcal I}_{AB} & = \bigg[-2\s\cg \beta\s \cg u_A^j\s %\left(\frac{\cg \beta}{\pi}\right)^{3/2} e^{-\cg \beta \cg u_A^2} \cg  n_A 
   \cg f_A^{(0)}\bigg] (-2\s\cg \beta) (\cg u_A - \cg u_B)^i \left(\frac{\cg \beta}{\pi}\right)^{3}  \cg{n}_B I^{ij} = \frac{\partial \cg f_A^{(0)}}{\partial \cg u_A^j} (-2\s\cg \beta) (\cg u_A - \cg u_B)^i\left(\frac{\cg \beta}{\pi}\right)^{3}  \cg{n}_B I^{ij}\,,
   \label{eq:CGAB}
\end{align}
where
\begin{align}
  I^{ij} & =\int \diff\Omega \int \diff^3  v_A\int  \diff^3 v_B\s  v_A^{j }\s  ( v_A^\prime-v^{\vphantom{\prime}}_A)^i \s e^{- \cg \beta (v_A^2 + v_B^2)} |\vec v_B - \vec v_A| \frac{\diff\sigma}{\diff\Omega} \notag \\[6pt]
    & = \frac{1}{3}\delta^{ij} \int \diff\Omega\int \diff^3 v_A\int  \diff^3  v_B \s v_A\cdot  (v_A^\prime - v^{\vphantom{\prime}}_A) \s e^{- \cg \beta (v_A^2 + v_B^2)} |\vec v_B - \vec v_A| \frac{\diff\sigma}{\diff\Omega} \notag\\[6pt]
    & = \frac{1}{3}\delta^{ij}\int \diff\Omega \int \diff^3 \g \int \diff^3 \G \left(\G - \frac{1}{2} \g\right)^i \left(-\frac{1}{2} \g^\prime + \frac{1}{2} \g^{\vphantom{\prime}}\right)^i \s \g \s e^{-\cg \beta(2 \G^2 + \g^2/2)}\frac{\diff\sigma}{\diff\Omega}\,.
\end{align}
In the last step, we have replaced the particle velocities by the relative and the center-of-mass velocities, \begin{equation}
    \vec{v}_{\rm rel}^{\s(\prime)}=\vec v_B^{\s(\prime)} - \vec v_A^{\s(\prime)}\,,\qquad \text{and} \qquad \vec \G =\frac{1}{2}\left(\vec v_A + \vec v_B\right)\,.
\end{equation}
For a spherically symmetric system, the integration along $\G_i$ vanishes. Therefore, we can show that the result depends only on the ``transfer cross section,''
   \begin{equation}
\sigma_T = \int \text{d}\cos{\theta} (1-\cos{\theta}) \frac{\text{d}\sigma}{\text{d}\cos{\theta}}\,.
\label{eq:sigT}
\end{equation}
This follows from
\begin{align}
    I^{ij} & =-\frac{1}{12}\delta^{ij}\int \diff\Omega \int \diff^3 \g \int \text{d}^3 \G \left(1- \frac{\g^\prime\cdot \g^{\vphantom{\prime}}}{\g^2}\right)\s \g^3\s e^{-\cg \beta(2 \G^2 + \g^2/2)}  \frac{\diff\sigma}{\diff\Omega}\notag\\[6pt]
   & = -\frac{(4\pi)^2}{12}\delta^{ij} \left[\int \diff \G \s e^{-2 \cg \beta \G^2} \G^2\right]\int \diff\Omega \int \diff \g \s e^{-\cg \beta \g^2/2} \g^5 \left(1-\cos{\theta}\right)\frac{\diff\sigma}{\diff\Omega}\notag \\[6pt]
   & = -\frac{\pi}{3} \left(\frac{\pi}{2\cg \beta}\right)^{3/2}\delta^{ij}\int \diff\Omega \int \diff \g \s e^{-\cg \beta \g^2/2} \s \g^5 \s \frac{\diff\sigma_T}{\diff\Omega}\,,
\end{align}
where the scattering angle is $\theta  =\arccos{(\vec v_{\rm rel}^{\s\prime} \cdot \vec v^{\vphantom{\prime}}_{\rm rel}/\g^2)}$.

Gathering all numerical factors and using \cref{eq:CGevolutionAppendix} to identify the effective force, we obtain:
\begin{align}
    \cg F_{AB}^i = \frac{\cg u_B^i - \cg u_A^i}{3 \sqrt {2\pi}} \cg {\rho}_B\s \cg \beta^{5/2} \int \diff\Omega \int \diff \g\s e^{-\cg \beta \s \g^2/2}\s \g^5 \frac{\diff\sigma_T}{\diff\Omega}\,,
\end{align}
which we rewrite below in terms of the dimensionless velocity $\g^*=\g\s \sqrt{\cg \beta}$\s, 
\begin{align}
    \cg F_{AB}^i & = \cg {\rho}_B\frac{\cg u_B^i - \cg u_A^i}{3 \sqrt {2\pi}}  \cg \beta^{-1/2} \int \diff\Omega \int \diff  \g^* \s  e^{- \g^{*\, 2}/2} \s \g^{*\,5} \frac{\diff\sigma_T }{\diff\Omega}\notag\\[4pt]
    &  \equiv \cg {\rho}_B\frac{\cg u_B^i - \cg u_A^i}{\cg \beta^{1/2}}  \frac{8\sqrt{2\pi}}{3}   \int \text{d}\s\!\cos{\theta}\frac{\text{d} \cg \sigma_T}{\text{d}\cos{\theta}}\,,
    \label{eq:FeffWlinear}
\end{align}
where the effective {transfer} cross section is 
\begin{equation}
    \frac{\text{d} \cg \sigma_T}{\text{d}\cos{\theta}} = \frac{\int \text{d}\g^* \s e^{- \g^{*\, 2}/2} \s \g^{*\,5} \frac{\text{d}\sigma_T}{\text{d}\cos{\theta}} }{\int \text{d}\g^* \s e^{- \g^{*\, 2}/2} \s \g^{*\,5}}\,.
    \label{eq:xs}
\end{equation}
This effective cross section provides a~\textit{universal} mapping between a velocity-dependent particle physics cross section and a constant macro-particle cross section, defined at a given $\cg \beta$. This result follows from the fact that one is not allowed to select just one velocity in the collision integral to coarse grain a single one-to-one collision among particles in different macro-particles.

Importantly,~\cref{eq:FeffWlinear} also shows that the velocity kernel we chose preserves momentum conservation at the macro-particle level, since the mass densities of macro-particles are assumed to be equal in the simulation. For example, this would not hold if we took the velocity kernel to be of the form $\delta^3 (\v v_A - \vcg u_A)$.

Finally, we show how the coarse-graining scheme acts on the gravitational force. Even though this force has been neglected in our previous derivation,  we show here that it can also be taken into account together with the effective force modeling particle collisions.
Since the gravitational force only depends on position, $\v F_{\rm gravity}\to {\vcg F}_{\rm gravity}$ trivially since we are assuming localized macro-particles in the sense of \cref{eq:GAdelta}. Therefore, all we need to show is that the derivative of the particle DF, which multiplies the gravitational force, maps to its coarse-grained counterpart:
\begin{align}
    \mathcal N\int \diff^3 v_A W_A \frac{\partial f_A^{(0)}}{\partial v_A^i} & = \int \diff^3 v_A W_A  (-2\s \cg \beta\s v_A^i) \left(\frac{\cg \beta}{\pi}\right)^{3/2} \s \mathcal N\s\cg n_A \s e^{-\cg \beta v_A^2} \notag\\[3pt]
    & = \int \diff^3 v_A (-2\s \cg \beta\s v_A^i v_A^j) (2\s\cg \beta\s \cg u_A^j) \left(\frac{\cg \beta}{\pi}\right)^{3}\s \mathcal N\s \cg n_A \s e^{-\cg \beta (v_A^2 + \cg u_A^2)}\nonumber \\[3pt]
    & = \frac{\partial \cg f_A^{(0)}}{\partial \cg u_A^i} \int \diff^3 v_A \frac{2\s\cg \beta\s v_A^2}{3} \left(\frac{\cg \beta}{\pi}\right)^{3/2} e^{-\cg \beta\s v_A^2} 
    =\frac{\partial \cg f_A^{(0)}}{\partial \cg u^i_A} \,.
    \label{eq:CGgravity}
\end{align}
Having shown that both the gravitational and self interactions of DM can be described by a coarse-grained force, it is possible to study these two effects together in a simulation by simply including the vector sum of these two contributions. Importantly, the proof that the coarse-graining scheme  trivially maps the gravitational force is relevant beyond the application of our framework to SIDM. It justifies why the same force can be applicable to both particles and macro-particles, even in collisionless DM simulations.

\clearpage

\bibliography{references}{}
\bibliographystyle{aasjournal}

\end{document}